# Does Cancer Growth *Depend* on Surface Extension?

Thomas S. Deisboeck, M.D. [1*], Caterina Guiot, Ph.D. [2,3],

Pier Paolo Delsanto, Ph.D. [3,4] and Nicola Pugno, Ph.D.[5]

[1] Complex Biosystems Modeling Laboratory, Harvard-MIT (HST) Athinoula A. Martinos Center for Biomedical Imaging, Charlestown MA 02129, USA; [2] Dip. Neuroscience, Università di Torino, Italy; [3] CNISM, Sezioni di Torino Università e Politecnico, Italy; [4] Dip Fisica and [5] Dip. Ingegneria Strutturale, Politecnico di Torino, Italy.

**\*Corresponding Author:**

Thomas S. Deisboeck, M.D.

Complex Biosystems Modeling Laboratory

Harvard-MIT (HST) Martinos Center for Biomedical Imaging

Massachusetts General Hospital-East, 2301

Bldg. 149, 13th Street

Charlestown, MA 02129

USA

Tel:    617-724-1845

Fax:    617-726-7422

Email: deisboec@helix.mgh.harvard.edu

**- Letter to the Editor -**

We argue that volumetric growth dynamics of a solid cancer *depend* on the tumor systems' overall surface extension. While this at first may seem evident, to our knowledge, so far no theoretical argument has been presented explaining this relationship explicitly. In here, we therefore develop a conceptual framework based on the universal scaling law and then support our conjecture through evaluation with experimental data.

First, let us consider that from a mechanical perspective any solid malignant tumor grows inside a given organ structure, i.e. is confined to a limited host volume. For instance, in the brain, surrounding bony skull defines this boundary, elsewhere it may be a less rigid organ capsule. Thus, a growing tumor should rapidly induce the build-up of a mechanical pressure *P*, which, as soon as all 'reserve rooms' have been used up, will increase sharply and may act as a growth constraint for the tumor. Without further adjustment, the tumor's growth curve would saturate at this point. Arguably, this phase coincides with reaching a critical cell density, $\rho$, and denotes the time point when tumor cell invasion starts, as (i) it offers a way to release 'excess' cells, reduce local cell density ($\rho$) and hence keep *P* manageable; (ii) since it reduces the adjacent tissue consistency through the cells' enzymatic activity or proteolysis, meaning that these invasive cells can reduce *P* directly (Deisboeck et al., 2005) or (iii) *increase* the total tumor surface A, in an effort to control *P*.

Secondly, tumor growth depends on nutrient and oxygen availability. As such, when dimensions exceed a few mm$^3$ (Folkmann, 1971), angiogenetic mechanisms are activated to provide the tumor with a nutrient supply system. Prior and/or in addition to angiogenesis surface-diffusion mechanisms are operating as well. The efficiency of the resulting nutrient supply system is related to its geometry. Up until now, attention has been devoted only to the dimensionality of the distributive system. Based on the assumption of a fractal vascular network, our extension of West's law (2001, 2004) to tumors conjectures that the rate of input energy $B$ is related to the mass $m$ by a power law of the type $B \propto m^p$ (Guiot et al., 2003). While in the case of a spherical tumor nourished only by diffusion, the scaling exponent $p$ should indeed correspond to 2/3, in the case of a distributive capillary network, however, West argues that $p = ¾$. Correspondingly, Banavar et al. (1999) found the same $p$ value in an effort to solve the problem of determining the exponent for a general distributive system. They claimed that $B$ is expected to scale as $M^{D/(1+D)}$ if the efficiency of the vascular network is maximized. (D is the dimensionality of the embedding space). Actually, in tumors it has been shown (see e.g. Baish and Jain, 2000) that the fractal dimension $\overline{D}$ of the vascularity ranges between 2 and 3 and, correspondingly, the value of $p = \overline{D}/3$ varies between 2/3 and 1. Also, both experimental (e.g., Carmeliet and Jain, 2000; Gazit et al., 1997) and theoretical considerations (Baish et al., 1996; Guiot et al, 2005; Guiot et al., In press) predict that $p$ varies at the onset of angiogenesis.

Now, Carpinteri and Pugno (2002) have developed universal scaling laws for energy dissipation in a different context, that is during the fragmentation of solids, by assuming a self-similar (i.e., fractal) size distribution of fragments. Their assumption implies a power law such as $N \propto r^{-\overline{D}}$, where $N$ is the number of fragments with size larger than $r$, and $\overline{D}$ is the so-called fractal exponent (a real positive number) of the fragment size distribution. Accordingly, they obtain by integration the total surface, $S$, of the fragments, as a function of their total volume, $V$,

Deisboeck T.S., et al.: "Does Cancer Growth *Depend* on Surface Extension?"as $S \propto V^{\overline{D}/3}$, with $2 < \overline{D} < 3$. In this case the value of the parameter $p$ would be related more to the system topology than to the occurrence of a distributive network. Returning then to our initial hypothesis that tumor growth depends on surface extension, we conjecture that such a model could be applied *also* to the case of a non- or pre-vascularized tumor that exhibits invasion into the surrounding tissue. In particular, at least in the early phases of invasion, a multicellular tumor spheroid (MTS) may develop an invasive branching structure nourished by diffusion (see Deisboeck et al., 2001; Gordon et al., 2003). For instance, let us consider a MTS, with radius $r$, that starts invading its microenvironment by generating branching structures that are composed of mobile cancer cells. Let each branch's mean radius be $\varepsilon r$, with $\varepsilon << 1$, whereas its length, $l$, varies in time. In order to limit the number of parameters, let us further consider the case in which the MTS surface is fully covered by the bases of the sprouting branches (for schematic, see **Figure 1a**). Then the number of branches is $N = 4/\varepsilon^2$ and the overall volume, $V$, becomes:

$$V = 4\pi(r^3/3 + l r^2), \qquad (1)$$

and the total surface:

$$S = 4\pi(r^2 + 2 l r/\varepsilon) \qquad (2)$$

Keeping a fixed $\varepsilon$ and $l$, we investigate, for different values of $r$, the relationship:

$$\ln(S) = k + p \ln(V) \qquad (3)$$

According to previous works reported in Deisboeck et al. (2001) and Gordon et al. (2003), both the MTS radius and length, *l*, of the invasive branches at different phases of invasion can be accurately evaluated up to the sixth day of *in vitro* culture. By inserting the corresponding data in Eq. (1,2,3) and assuming reasonable values for the parameter $\varepsilon$, the '*p*' value which best fits the MTS growth can be evaluated (see **Figure 1b**). For *l* = 0, i.e. before branching starts, it is obviously *p* = 2/3, but, as *l* increases, *p* increases as well, reaching the value of 0.82, which is scarcely affected by the choice of $\varepsilon$ (between 0.1 and 0.001).

**Figure 1 here**

If the invasive branching process would continue, the same mathematical model would predict a further increase in *p* if l continues growing. In particular, assuming $\varepsilon$ = 0.001, it will yield a *p* value very close to 1 when l >> $r_{max}$. In conclusion, a branch-inducing MTS may indeed yield *p* values in the range (2/3; 1), depending on the length or extent of its invasive architecture - *even in the absence* of angiogenesis (which by itself has been credited for moving *p* to ¾). This point is rather critical since it argues that invasion-mediated tumor surface expansion, and thus surface-diffusion nourishment can and should *precede* neovascularization, which may take more time in order to become effective.

In summary, as an amendment to a recent paper of ours (Guiot et al., In press), which claims that a transition occurs for *p* from 2/3 to 1 for an angiogenesis-dominant nutrition supply mechanism, we argue here that, even before the onset of angiogenesis, i.e. in the early phase of tumor cell invasion, the parameter *p* can *vary* according to changes in the topological configuration.

Therefore, we conjecture that cancer system growth, at least in its early stages, *does* critically depend on surface extension and thus on rapid tissue infiltration. Cautiously extrapolated to the experimental research side, our conjecture may yield intriguing insights into the sequence of events involved in the molecular progression pathways. For instance, one could hypothesize that genetic and epigenetic profiles which increase the tumor's invasive behavior are selected for at a relatively early stage – a process that would have implications both for diagnostics and therapeutics alike. Furthermore, we argue that this surface-expansion due to nourishment requirements complements the cell density- and thus largely mechanically-driven trigger, described in detail in Deisboeck et al. (2005).

Lastly, in the presence of vasculature, the aforementioned mechanism may be responsible for regional heterogeneities in the prevailing nourishment process at the same time point, in that wherever the surface to volume ratio is favorably high, a diffusion-dominant supply mechanisms is sufficient (i.e., in the regions that show extensive cell invasion branching), whereas for tightly packed volumetric objects such as the main tumor growth core, for microsatellites and metastases, angiogenesis, and therefore $p = 3/4$, is the desired, necessary nutrient-supply mechanism. Even the occurrence of *p* values larger than ¾, as observed by Guiot et al. (In press) can then be explained by such a mechanism.

This letter therefore presents further evidence that many different tumor growth conditions can be described with a relatively simple law such as the one proposed by West et al., provided the scaling parameter *p* is kept variable in space and time to account for different nourishment conditions.

**Acknowledgements:** This work has been supported in part by NIH grants CA 085139 and CA 113004 and by the Harvard-MIT (HST) Athinoula A. Martinos Center for Biomedical Imaging and the Department of Radiology at Massachusetts General Hospital.

**Figure & Caption:**

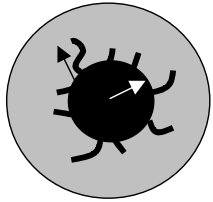
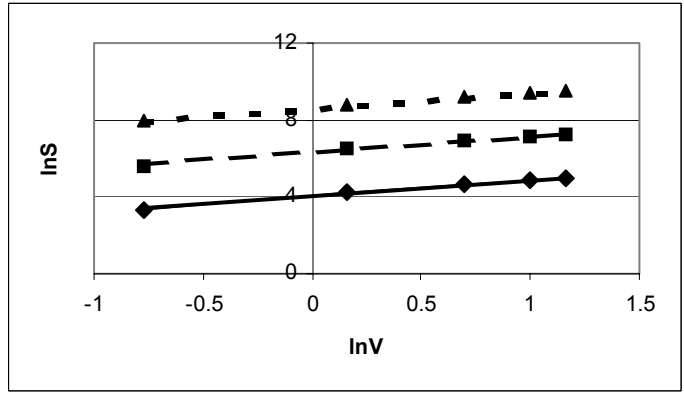

a)                                            b)

**Figure 1:** **(a)** The white arrow marks the radius *r* and the black arrow the length *l* used to compute *V* and *S*. **(b)** '*p*' is estimated as the linear coefficient in Eq.(3), and *lnS* and *lnV* are taken from experimental data (Deisboeck et al., 2001; Gordon et al., 2003); $\varepsilon$ is the only free parameter. Simulations with $\varepsilon = 0.1$ (solid line), $\varepsilon = 0.01$ (dotted line) and $\varepsilon = 0.001$ (dotted line) yield *p* = 0.818, 0.826 and 0.827, respectively ($R^2 > 0.99$ in all cases).